\documentstyle[12pt,fleqn,epsfig]{article}
\font\fontb=cmr12 scaled\magstep2 
\font\fontc=cmr12 scaled\magstep1 
\def\beq {\begin{eqnarray}} 
\def\eeq {\end{eqnarray}} 
\def\beqn {\begin{eqnarray*}} 
\def\eeqn {\end{eqnarray*}} 
\def\neqn {\nonumber}

\def\ni {\noindent} 
\def\new {\newpage} 
\def\PL #1 #2 #3 {Phys. Lett.~{\bf#1} (#2) #3} 
\def\NP #1 #2 #3 {Nucl. Phys.~{\bf#1} (#2) #3} 
\def\ZP #1 #2 #3 {Z.~Phys.~{\bf#1} (#2) #3} 
\def\PR #1 #2 #3 {Phys. Rev.~{\bf#1} (#2) #3} 
\def\PP #1 #2 #3 {Phys. Rep.~{\bf#1} (#2) #3} 
\def\PRL #1 #2 #3 {Phys. Rev.~Lett.~{\bf#1} (#2) #3} 
\def\PTP #1 #2 #3 {Prog. Theor.~Phys.~{\bf#1} (#2) #3} 
\def\MPL #1 #2 #3 {Mod. Phys.~Lett.~{\bf#1} (#2) #3} 
\def\IJM #1 #2 #3 {Int. J.~Mod.~Phys.~{\bf#1} (#2) #3} 
 
\def\ra {\rightarrow} 
\def\etal {{\it et al}.} 
\def\eg {{\it e.g}.} 
\def\ie {{\it i.e}.}

\def\bra {\langle} 
\def\ket {\rangle} 
\def\GeV{\mbox{GeV}} 
\def\MeV{\mbox{MeV}}

\def\sin {\mbox{sin}}

\begin{document} 
\begin{titlepage} 
\vspace* {-1cm} 
\baselineskip=1cm 
\begin{center}{\fontb Chiral-odd transversity spin structure function  
$h_1(x)$ of the nucleon in a constituent quark model} 
 
\baselineskip=0.75cm 
\vspace {1cm} 
{\fontc  
Katsuhiko Suzuki\footnote{Alexander von Humboldt fellow, e-mail address :  
                ksuzuki@physik.tu-muenchen.de}} 
 
{\em Institut f\"{u}r Theoretische Physik,  
Technische Universit\"{a}t M\"{u}nchen}\\ 
{\em D-85747 Garching, Germany}\\ 
\vskip 0.2cm 
 
and 
 
\vskip 0.2cm 
{\fontc Takayuki Shigetani\footnote{e-mail address :  
shige@postman.riken.go.jp}}

{\em    Computation Center,  
        The Institute of Physical and Chemical Research (RIKEN)}\\ 
{\em Wako, Saitama 351-01, Japan}\\

\vspace {1cm}

{\bf Abstract} 

\end{center} 

\baselineskip=0.7cm 
 
\ni 
We study the chiral-odd transversity spin-dependent quark distribution  
function  $h_1 (x)$ of the nucleon in a constituent quark model.    
The twist-2 structure functions, $f_1(x)$, $g_1(x)$ and $h_1(x)$  
are calculated within the diquark spectator approximation.   
Whereas an inequality $f_1(x) > h_1(x) > g_1(x)$ holds with the interaction 
between quark and diquark being scalar,  
the axial-vector effective quark-diquark interaction, which contributes to 
the  $d$-quark distribution, does not lead to such a simple relation.   
We find that $h_1(x)$ for the $d$-quark becomes somewhat smaller than  
$g_1^d (x)$, when we fix the model 
parameter to reproduce known other structure functions.   
We also include corrections due to the non-trivial structure of the 
constituent quark, which is modeled by the Goldstone boson dressing.   
This improves agreements of $f_1(x)$ and $g_1(x)$ with experiments, and  
brings further reduction of $h_1^d(x)$ distribution.   
Consequences for semi-inclusive experiments are also discussed.  

\vspace {0.4cm}  

Key Words: structure function, nucleon spin, constituent 
quark model, chiral symmetry

\end{titlepage} 
\baselineskip=0.7cm 
\new
\ni 
{\bf 1 Introduction}

Study of the nucleon spin structure triggered by the unexpected EMC 
result \cite{EMC}  
becomes one of the most challenging issues in the hadron physics.   
The spin structure function obtained by the 
polarized lepton-hadron deep inelastic scattering tells us a lack of  
our understandings of the quark and gluon dynamics in the nucleon.   
>From the lepton-hadron deep inelastic scattering, one can obtain the  
longitudinal spin structure function $g_1 (x)$\cite{Exp_g1}   
and the transverse spin structure function,  
$g_2(x)$, which contains information on twist-3 pieces\cite{Exp_g2}.   
These structure functions reveal the helicity distribution function 
as well as the quark-gluon correlation in the nucleon\cite{Spin_review}.   
Further approach using the semi-inclusive process is now planed,  
which makes it possible to know detailed decomposition of the nucleon spin 
structure.

Recently, chiral-odd structure function is discussed as complementary tool  
to study the structure of the nucleon\cite{Soper,JaffeJi}.    
In the lepton-hadron  
scattering, in which the chirality is conserved, only the chiral-even  
structure functions can be observed.   
On the other hand, in the Drell-Yan process, the chirality changing  
process is possible, and thus the chiral-odd structure function  
is expected to be measured.

In terms of the quark-quark density matrix, the twist-2 structure  
functions of the nucleon are expressed as\cite{JaffeJi}, 
\beq 
\int_{}^{} {{{d\lambda } \over {2\pi }}} 
e^{i\lambda x}\left\langle {P,S|\psi (0)\gamma _\mu \psi 
 (\lambda n)|P,S} \right\rangle =2\left[ {f_1(x)+ \mbox{h.t.}} \right] 
\eeq 
\beq 
\int_{}^{} {{{d\lambda } \over {2\pi }}} 
e^{i\lambda x}\left\langle {P,S|\psi (0)\gamma _\mu \gamma ^5\psi  
(\lambda n)|P,S} \right\rangle =2 
\left[ {g_1(x)P_\mu (S\cdot n)+\mbox{h.t.} } \right] 
\eeq 
\beq 
\int_{}^{} {{{d\lambda } \over {2\pi }}} 
e^{i\lambda x}\left\langle {P,S|\psi (0)\sigma _{\mu \nu } 
i\gamma ^5\psi (\lambda n)|P,S} \right\rangle = 
2\left[ {h_1(x)(S_{\bot \mu }P_\nu -S_{\bot \nu }P_\mu )+\mbox{h.t.}} \right] 
\eeq 
(h.t.~denotes the higher twist contributions.) 

\ni 
Here, we followed the definition of Ref.~\cite{JaffeJi}.   
$f_1(x)$ is the usual spin-independent quark distribution,  
and $g_1(x)$ the quark helicity distribution function in the  
longitudinally polarized nucleon.  These two distribution functions  
are chiral-even,  
and can be measured by the lepton-hadron scattering.   
The {\em transversity } spin distribution function $h_1(x)$ is chiral-odd, in  
which we are mainly interested in this paper.   
The $h_1(x)$ structure function corresponds to a target helicity-flip 
amplitude in the helicity basis, and thus does not have simple partonic 
probabilistic  
interpretation.  As pointed out in Ref.~\cite{JaffeJi}, however,  
the transversity spin distribution can be understood as difference of  
numbers of valence quarks with eigenvalues $+1$ and $-1$ of the transverse 
Pauli-Lubanski operator in the transversely polarized nucleon.

Our aim of this paper is to present $h_1(x)$ using low energy effective quark 
model, which is already used to calculate known structure functions  
$f_1(x)$ and $g_1(x)$.   In particular, we discuss the flavor 
dependence of the 
transversity spin distribution $h_1(x)$ in detail, compared with other twist-2 
quark distributions.     
We evaluate quark distribution functions $f_1(x)$, $g_1(x)$ and $h_1(x)$  
of the nucleon based on a constituent quark picture, which is successful  
to describe the low energy hadron properties.     
Here, current quarks are assumed to acquire their dynamical masses due to  
spontaneous breakdown of the chiral symmetry following to   
the Nambu and Jona-Lasinio model\cite{NJL}.      
We take the model developed in Ref.~\cite{MM,Suzuki_diq,Kulagin} to  
calculate the structure 
functions, where the nucleon is described as a bound state 
of the quark and diquark.

For effective interaction between quark and diquark,  
we take both scalar and axial-vector types from the symmetry 
consideration.    
In principle, we could solve a bound state equation of the  
quark-diquark for the nucleon.  However, it needs difficult procedure,  
and is first recently done by Kusaka {\etal} for only the scalar  
channel\cite{Kusaka}.   
Here, we simply assume reasonable forms of the quark-diquark vertex  
functions for the scalar and 
axial-vector channels so as to reproduce the observed structure function  
$f_1(x)$ and $g_1(x)$, and try to describe $h_1(x)$ structure function.

There are several works to investigate the transversity spin distribution  
function $h_1(x)$ within the quark models\cite{JaffeJi,Ioffe,CDM,QSR}.   
Since $h_1(x)$ is defined by  the chirality-violating operator,  
the transversity spin structure function may depend on the chiral-odd  
operators of the theory, {\eg}~the quark condensate of the QCD  
vacuum\cite{Ioffe}.

Recently, Soffer has proposed that such twist-2 quark distribution functions,  
$f_1(x)$, $g_1(x)$ and $h_1(x)$, are subject to   
{\em Soffer's inequality}, which is derived from a general 
argument\cite{Soffer,Soffer2}.    
\beq 
f_1(x)  + g_1(x)  \geq  2 | h_1(x)| 
\label{soffer1} 
\eeq 
We discuss behavior of this inequality in our model.

In particular, the simple relativistic quark model such as the bag 
model\cite{JaffeJi},  where the   
relativistic quarks independently move inside 
the confinement region, leads to a relation; 
\beq 
 f_1(x) + g_1(x) =   2 | h_1(x)|  
\label{bag1} 
\eeq 
\beq 
| h_1(x)| >   | g_1(x) | 
\label{bag2} 
\eeq 
The first equation (\ref{bag1}) means a saturation of the Soffer's inequality 
in the model, discussed by Goldstein {\etal}\cite{Soffer2}.   
The second inequality comes from roles of the lower component of the Dirac 
spinor.   
Indeed, 1st moments of $g_1(x)$ and $h_1(x)$, axial charge and tensor 
charge,  are given by 
\beqn 
\Delta q  &=& \int [dr] \left[ F^2 - \frac{1}{3} G^2 \right]\\ 
\delta q  &=& \int [dr] \left[ F^2 + \frac{1}{3} G^2 \right] 
\eeqn 
where $F$ and $G$ are upper and lower radial wave functions of the Dirac 
spinor.   Eq.~(\ref{bag2}) is a universal relation of the relativistic quark 
potential model independent of the model parameters.

However, recent lattice QCD simulation indicates  
the 1-st moment of the $| h_1(x)|$ is larger than that of  
$ | g_1(x) |$ for the $d$-quark distribution function\cite{h1lattice}.   
In addition,  
the QCD sum rule study suggests a very small value of the $d$-quark 
tensor charge\cite{QSR}.

In the present study, we take both scalar and axial-vector  
quark-diquark-nucleon vertices, and examine such relations.   
When the nucleon is a bound state of the quark and the scalar diquark,  
we exactly obtain $f_1(x) + g_1(x) =  2 h_1(x) $ and    
$h_1(x) > g_1(x)$\cite{Artru}, 
which is quite similar with the bag model calculations.     
However, for the axial-vector vertex case, which determines the $d$-quark 
distribution function, such simple relations are not maintained.   
We find $| h_1^d (x)| < | g_1^d(x) | $ or they are comparable at least,  
if we fix the model parameters to reproduce the other structure functions.

We also study corrections to the quark distributions from dressed structure of 
the constituent quark.   
Constituent quarks are assumed to be quasi-particles of the QCD vacuum, and  
have non-trivial structure.   
We introduce the Goldstone boson cloud, $\pi,K,\eta$,   
around the constituent quark studied in Ref.~\cite{Suzuki}.   
Such a dressing produces crucial effects on the structure function, namely, 
reduces a probability to find a bare quark state and also changes the spin 
structure by emitting the GS boson to relative $P$-wave state.    
Combining these contributions with the results of quark-diquark model, we  
show that shape of the transversity spin distributions are considerably 
different from those of the helicity distributions.

This paper is organized as follows.  In Section 2,  
we calculate the chiral-odd type forward  
scattering amplitude introduced in Ref.~\cite{Ioffe}, which provides the  
transversity spin distribution $h_1(x)$.    
Relations of three structure function are examined in some detail.   
In Section 3, numerical results are given performing the $Q^2$ evolution  
of the distribution functions.  $f_1(x)$ and $g_1(x)$ are 
compared with the experimental data, and $h_1(x)$ is presented for forthcoming 
experiment.   
We introduce the dressed structure of the constituent structure due to the  
Goldstone boson cloud, developed recently by Suzuki and  
Weise\cite{Suzuki}.    
Inclusion of the GS boson dressing improves the agreements substantially.   
Final section is devoted to summary and discussions.

 
\vspace{2cm} 
 
\ni 
{\bf 2 Calculations of the twist-2 quark distribution functions}

Purpose of the present paper is to study the structure function measured  
at high energies by means of the 
low energy quark model of the nucleon.  Basic procedure to connect  
calculations of the quark model with the high energy experimental data is 
summarized as follows \cite{Jaffe};     
We evaluate twist-2 matrix elements of the structure function within the 
effective quark model at the scale $\mu \sim 1 \GeV$, where the effective   
models are supposed to work.    
Calculated structure function has no physical meaning at 
this scale $\mu$, but plays a role of the boundary condition for the QCD 
evolution equation.  With the help of the perturbative QCD, the results are 
evolved from the low energy model scale $\mu$ to the high momentum scale,  
at which the experimental data exist.   
Higher twist contributions can be neglected, 
since we are interested in the Bjorken limit $Q^2 \ra \infty$.   
Comparison of the model calculations with the experimental data  
reveals non-perturbative dynamics in the deep inelastic experiments.

Let us start calculating nucleon structure functions, $f_1(x)$, $g_1(x)$ and  
 $h_1(x)$.   
We adopt the phenomenological quark-diquark model for the 
nucleon \cite{MM,Suzuki_diq,Kulagin}.    
In this approach, the virtual photon-nucleon forward scattering amplitude is  
illustrated in Fig.1.   
Here, the constituent quark is struck out by the  
high momentum virtual photon with a  residual diquark being spectator.   
We treat diquarks as only the spectators, though the structure  
of the diquark itself might be crucial when we go beyond the 
diquark spectator approximation \cite{Suzuki_diq}.   
Regarding the nucleon-quark-diquark effective vertex, we deal with both scalar 
and 
axial-vector types, which are suggested  by the success of the $SU(6)$ 
spin-flavor symmetry approach. 
\beq 
V _{S} &=& {\bf 1} \phi_S(p^2) \\ 
V_{V}^\mu &=& \gamma^\mu \gamma^5 \phi_V (p^2) 
\label{vertex} 
\eeq 
where $\phi_S (p^2)$ and $\phi_V(p^2)$ are momentum cutoff functions, 
specified  later.

\begin{figure} 
\begin{center} 
\psfig{file=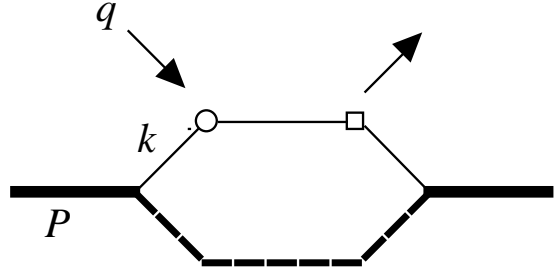,height=1.5in} 
\caption{The forward scattering amplitude in the quark-diquark model 
\label{fig1}}   
\end{center} 
\vspace{-0.2cm} 
The thick and thin solid lines denote the nucleon and the quark,  
respectively.  The spectator 
diquark is depicted by the dashed line.  The circle and box indicate  
the external currents, which are 
the axial-vector and scalar currents for $h_1(x)$.    
For $f_1(x)$ and $g_1(x)$ cases the current probes the nucleon structure as
the vector and axial-vector, respectively.  
\end{figure} 

In the following studies, we use standard definitions for the quark 
distribution functions\cite{JaffeJi}, 
\beqn 
F_1(x) = \frac {1}{2} e_a^2 \left[ f_1^a (x) + f_1^{\bar a} (x) \right] 
\eeqn 
where $F_1(x)$ is the standard electro-production structure function,  
and $e_a$ is the electric charge of the quark.  
Summation  $a$ is over quarks for all flavors.

At first, we show basic steps to derive the spin independent distribution  
$f_1(x)$ for the scalar channel.   
Explicit expressions for $f_1(x)$ and $g_1(x)$ are already given in  
literature\cite{MM,Kulagin}.   
We define the virtual photon-nucleon forward scattering amplitude as, 
\beq 
T_{\mu \nu }(q,P,s)=i\int {{{d^4\xi } \over {(2\pi )^4}} 
e^{iq\cdot \xi }\left\langle {P,s|TJ_\mu (\xi) \, J_\nu (0)  
|P,s} \right\rangle } 
\eeq 
where $J_\mu$ is the vector current.   
Here, $q$ is the momentum of the virtual photon, and $p$ the nucleon  
momentum with the mass $P^2 = M^2$.  $s$ is the nucleon spin, which    
satisfies  $p \cdot s = 0$ and $s^2 = -1$.   
We also introduce variables, $\nu = P \cdot q / M $ and the  
Bjorken-$x$ $x = -q^2 / 2M \nu = Q^2 / 2 M \nu$.   
By virtue of the optical theorem, this forward scattering amplitude is related
with the hadronic tensor, which provides the structure function.

In this model, the matrix element is calculated as, 
\beq 
\hspace{-0.5cm}T_{\mu \nu} = i  
\int_{}^{} {{{d^4p} \over {(2\pi )^4}}} 
\frac{1}{2} \mbox{Tr}\left[ V_i (k)S(k)\gamma _\mu S(k+q)\gamma _\nu  
S(k) V_i (k) D_i (k_2) (\not P+M)(1+\gamma ^5 \not s) \right] 
\label{ampl_f}
\eeq 
\ni 
where  
momenta delivered by the quark and diquark are $k$ and $k_2 = P - k$, 
respectively.    
Here, the quark propagator $S(p)$ with the constituent quark mass $m$  
is written as, 
\beq 
S(p)={1 \over {\not p-m}} 
\label{pro-cq} 
\eeq 
The scalar (spin-0) diquark propagator is given by,  
\beq 
D_S (p) = \frac{1} {p^2 - m_D^2} \,\, , 
\label{pro-sdq} 
\eeq 
and for the axial-vector (spin-1) diquark 
\beq 
D_V (p)_{\mu \nu} = \frac{1} {p^2 - m_D^2}  \left(- g_{\mu \nu} +  
\frac {p_{\mu} p_{\nu} } {p^2} \right) 
\label{pro-vdq} 
\eeq 
Here, $m_D$ is masses of the diquarks.   
 
Using the identity, 
\beqn 
\gamma _\mu \not q\gamma _\nu =S_{\mu \rho \nu \sigma } 
q^\rho \gamma ^\sigma -i\varepsilon _{\mu \nu \rho \sigma } 
q^\rho \gamma ^\sigma \gamma ^5 \;\;, 
\eeqn 
\ni 
one can get the spin independent and dependent distribution functions  
after suitable projection\cite{Spin_review}.    
For instance, using the projector ${\cal P}_{\alpha \beta} = \frac{1}{4} 
\left[ \frac{ 2 x } {P \cdot q} P_\alpha P_{\beta} - g_{\alpha \beta} 
\right]$, we can get the unpolarized quark distribution $f_1(x)$.

We introduce the light-cone variable, $x^{\pm} \equiv x^0 \pm x^3$.  
Non-vanishing contribution to the imaginary part of the forward scattering
amplitude arises from poles of the struck quark propagator $[(k+q)^2-m^2]$ 
and the spectator diquark propagator $[k_2^2 - m_D^2]$.  
The former is reduced to a condition that the Bjorken-$x$ is identified with
the longitudinal momentum fraction of the parton.  
\beqn
\delta [(k+q)^2- m^2] = \frac{1}{2 P \cdot q} \delta \left[ \frac{k^+}{P^+} 
- x \right]
\eeqn
The latter one yields,
\beqn
\delta [k_2^2- m_D^2] = \delta [ (1-x) P^+ k_2^- - k_T^2 - m_D^2] \; .
\eeqn

Inserting the above results into Eq.~(\ref{ampl_f}), one finds
\begin{eqnarray} 
f_S (x)&=& \int {{{dk^+ dk^- d^2 k_T } \over {2 (2\pi )^4}}}{{\phi_S ^2}  
\over {(k^2-m^2)^2}}2\pi \delta \left[k^-_2 - \frac {k_T^2 + m_D^2}
{(1-x) P^+ } \right]
\nonumber\\ 
& &   \times \, {2\pi  \delta(k^+ - x P^+) 
\over {2 P \cdot q  }} 
\frac{1}{2}\mbox{Tr}[(k\kern -1.7mm /+m) q \kern -1.7mm / (k\kern -1.7mm /+m)
(P\kern -2.5mm /+M)  ]\nonumber\\ 
&=& \int^{k_{\max }^2}_{-\infty}  {{{dk^2} \over {8{\pi}^2}}} 
{{\phi_S ^2(k^2)} \over {(k ^2 - m^2)^2}} \neqn \\ 
&&\hspace{1cm} \times [x(M^2+2mM-m_D^2)+m^2 - (1-x)k^2] 
\label{fs0} 
\eeq 
where  
\begin{eqnarray} 
k_{\max}^{2} = - {x \over 1-x}m_D^2 + x M^2 \;\;  . 
\label{pin} 
\end{eqnarray} 
It is possible to rewrite Eq.~(\ref{fs0}) as 
\beq 
f_S (x)= \int^{k_{\max }^2}_{-\infty}  {{{dk^2} \over {8{\pi}^2}}} 
{{\phi_S ^2(k^2)} \over {(k ^2 - m^2)^2}}  [(x M + m) + k_T^2] 
\label{fs} 
\end{eqnarray} 
with the transverse momentum $k_T$ given by 
\beqn 
k_T^2 = (1 - x)x M^2 - x m_D^2 -( 1-x) k^2 
\eeqn 

Now we come back to the chiral-odd distribution function $h_1(x)$,  
which can not  
be probed by the electro-magnetic current, as mentioned in Section 1.    
Ioffe and  Khodjiamirian introduced the following scattering  
amplitude\cite{Ioffe},  
which allows us to evaluate  $h_1(x)$ as, 
\beq 
T_\mu (q,P,s)= i \int {d^4\xi } 
e^{iq\cdot \xi } {{1} \over {2}} \left \langle  
P,s|T \{ J_{\mu 5}(\xi ) \, J(0) +  J(\xi ) \, J_{\mu 5}(0)  \}
|P,s  \right\rangle  
\eeq 
where we choose the transverse nucleon spin $s=(0,1,0,0)$.   
This forward scattering amplitude is related with the $h_1 (x)$   
structure function.   
\beq 
T_\mu (q,P,s)&=&\left( {s_\mu -{{P \cdot q} \over {q^2}}q_\mu }  
\right) \tilde  h_1(x) \neqn \\ 
&&+\left( {P_\mu -{{P\cdot q} \over {q^2}}q_\mu }  
\right)(q\cdot s) \tilde  l_1(x)+\varepsilon _{\mu \nu \lambda \sigma } 
q_\lambda s_\sigma (q\cdot s) \tilde  l_2(x) 
\eeq 
Hence, one can obtain the transversity spin distribution function  through the 
optical theorem;   
\beq 
h_1(x) = - \frac{1}{\pi} \mbox{Im} \tilde h_1(x) 
\eeq 
Straightforward calculations yield  
the following expression for the scalar spin-0 diquark  
spectator process\cite{Artru}; 
\beq 
h_S (x)=  \int_{- \infty} ^{k^2_{\max} } 
\frac{dk^2}{8\pi^2} { {\phi_S (k^2)} \over {(k^2-m^2)^2} } 
\left[  (xM+m)^2  \right] 
\label{hs} 
\eeq 
where we use $(k \cdot s )^2 = k_x^2 = k_T^2 / 2$.   
It is interesting to note that resulting $h_S(x)$ is proportional to 
masses of the 
nucleon $M$ and the constituent quark $m$, which are the chiral symmetry 
breaking parameter of the model.   
If we assume a `un-realistic' chiral symmetric world, $M=m=0$, the $h_S (x)$  
distribution function vanishes.   This is different from  $f_1(x)$  
and $g_1(x)$, which are non-zero even if all the particles are mass-less.

Using (\ref{pro-cq}) and (\ref{pro-vdq}) and summing spin-1 diquark  
helicities, we also get the $h_1(x)$ for the axial-vector  
diquark spectator process.   
\beq 
h_V (x)&=&  \int_{-\infty}^{k^2_{\max}}  
\frac{dk^2}{(8\pi^2)}{{\phi_V (k^2)} \over {(k^2-m^2)^2}} 
\left[-x(M-m)^2   \right. \neqn \\ && \hspace{-1.5cm} \left. +  
{1 \over {m_D^2}}\left[ {(k^2-m^2)(k^2-M^2)+x(k^2-M^2)(m^2-M^2)- 
k_T^2(m+M)^2} \right] \right] 
\label{hv} 
\eeq 
It is not easy to see how  $h_1^d(x)$ behaves in the chiral mass-less 
limit.  Even if we take $m=M=0$,  a term $\sim k^4 / m_D^2$ 
survives and $h_V (x) \ne 0$.

Before constructing the nucleon structure function with the explicit  
spin-flavor wave function, we shall discuss relations of these quark  
distribution functions and the Soffer's inequality\cite{Soffer}.   
We show in Fig.2  $f_S(x)$, $g_S(x)$ and $h_S(x)$  
calculated by the quark-scalar diquark model.   
In this case, the relation $f_S (x)> h_S (x) > g_S (x) $ holds, 
independent of the model parameters, and the Soffer's inequality is  
saturated $f_S(x) +  g_S(x) = 2 h_S(x)$.  This behavior is similar with the 
bag model.

\begin{figure} 
\begin{center} 
\psfig{file=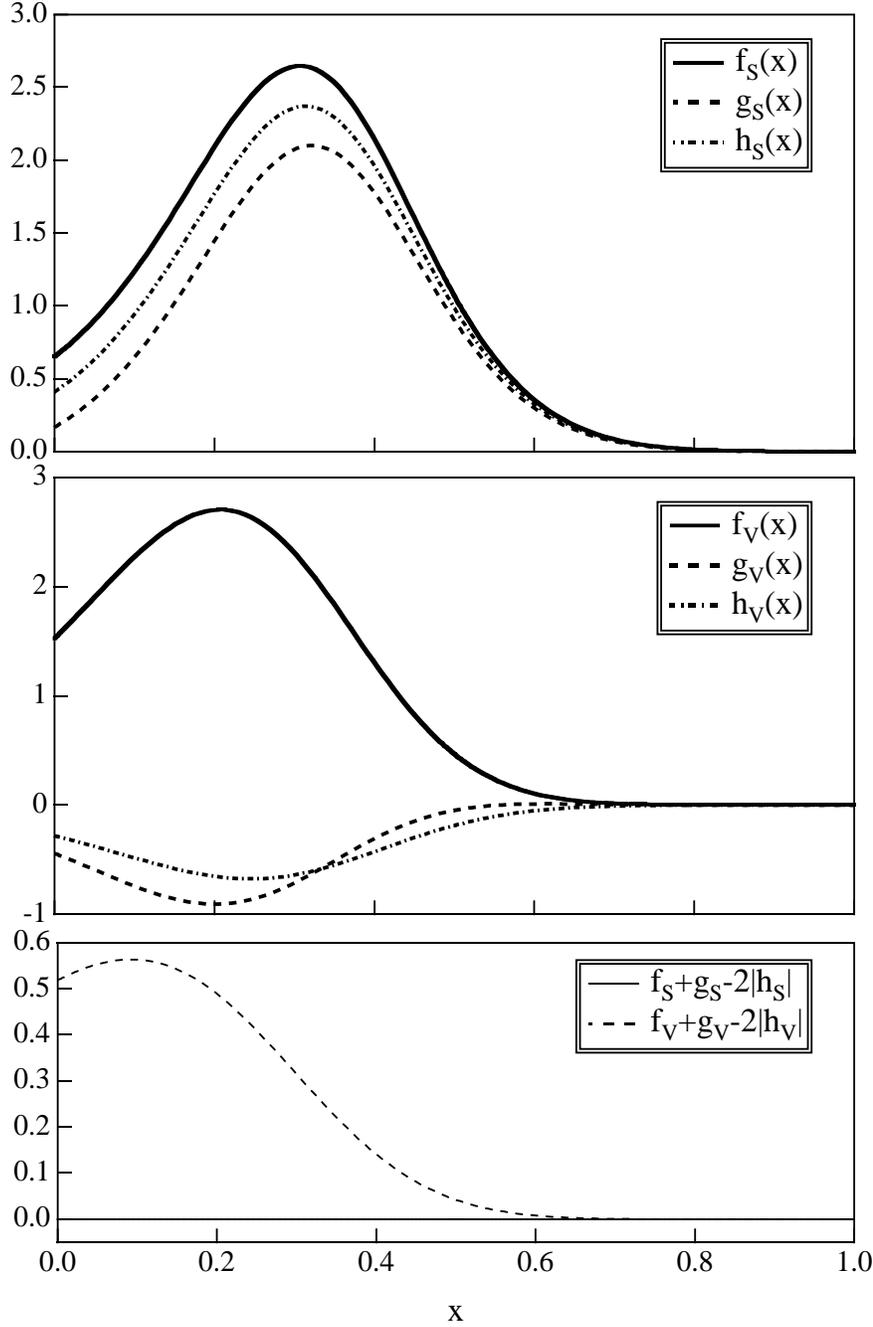,height=7in} 
\caption{Quark distribution functions in the quark-diquark model of the  
nucleon\label{fig_soffer}} 
\end{center} 
\vspace{-0.2cm} 
The upper 
figure shows the results of the spin-0 diquark spectator process (scalar  
channel), and the middle one  the results of the spin-1 diquark spectator  
(axial-vector channel).  In each figure, the quark distributions  
$f_1(x)$, $g_1(x)$ and $h_1(x)$ are shown by the solid, dashed and 
dash-dotted curves, respectively.   The lower figure shows  
$f_1(x) + g_1(x) - 2| h_1(x) |$ for the scalar and axial-vector channel  
by the solid and dashed curves, respectively.     
\end{figure} 

The spin-1 diquark spectator process shows rather non-trivial behavior.   
The quark distribution functions are shown in the middle of  
Fig.2.   
The Soffer's inequality is satisfied, but can not be saturated.   
Moreover, relative magnitude of $g_V(x)$ and $h_V(x)$ depends on the  
choice of the parameters.    
Here, we use the parameter set, which is suitable to reproduce other  
quantities such as the quark momentum fraction and $g_A$.   
We can not find any simple relation for $h_V(x)$ and $g_V(x)$.   
For completeness, we plot $f(x) + g(x) - 2|h(x)|$ in a lower part of  
Fig.2.   
In the spin-0 diquark spectator case, 
this quantity becomes exactly zero, which means  
a saturation of Soffer's inequality.   
Result of the spin-1 diquark case also satisfies a positive condition for  
 $f(x) + g(x) - 2|h(x)|$.

In terms of these distribution functions, we construct the nucleon structure 
functions with the spin and flavor structure.    
Since we have already included the spin structure of the diquarks to 
calculate the distribution functions and summed up over diquark helicities, 
we need only flavor structure of the nucleon wave
function, which is given by
%
%
\beq
|p >&=&  {1 \over {\sqrt {2}}} S(ud) u  + \frac{1}{\sqrt{6}} A(ud) u 
- \frac{1}{\sqrt{3}} A(uu) d \, , 
\eeq
where $u, d$ denote the up-, and down-quarks, and $S(ij)$ and $A(ij)$  
are the spin-0 and spin-1 diquarks, respectively.  Here we assume the $SU(4)$ 
spin-flavor symmetry.  
Quark distribution functions $q_1^u(x)$ and $q_1^d(x)$ with $q=f,g,h$ are 
expressed as 
\begin{eqnarray}
q_1^u (x)={3 \over 2}q_{S}(x)+{1 \over 2} q_{V}(x) \; ,
\label{uval} 
\end{eqnarray} 
\begin{eqnarray} 
q_1^d (x)= q_{V}(x)  \; .  
\label{dval} 
\end{eqnarray} 
Here $q_{S}$ and $q_{V}$ are corresponding quark distribution functions  
presented in Fig.2  
with the spin-0 and spin-1 diquark being spectators, respectively.

As easily seen in Eqs.~(\ref{uval},\ref{dval}), dominant contribution to the  
$u$-quark distribution comes from the spin-0 diquark spectator process, and  
the $d$-quark distribution is governed by the spin-1 diquark process.   
Recall that Eq.~(\ref{hs}) implies the transversity distribution vanishes 
when $m=M=0$, 
$h_1^u(x)$ distribution would become very small,  
if the chiral symmetry were not spontaneously broken.

Now we fix the model parameters with physically motivated constraints.   
The spin-0 diquark is expected to be tightly correlated quark-quark  
state due to the color spin-dependent interaction, as discussed in 
several works\cite{MM,Suzuki_diq}.   
In particular, experimental data of  
the non-leptonic hyperon week decay require the strong correlation in the  
spin-0 diquark, which resolves the long standing problem of the  
$\Delta I = 1/2 $ rule\cite{Stech,Suzuki_PG}.

Here, we take into account differences between the spin-0 and spin-1 
diquarks in the following ways.   
We assume that the mass  
difference of the diquarks is a few hundreds  MeV  
from a consideration of the $N$-$\Delta$ mass splitting.   
Effective size of the spin-1 diquark is expected to be larger than that 
of the  
spin-0 diquark, because it is assumed to be weakly bound state of quarks.   
We introduce the following parametrization for the quark-diquark vertex 
functions\cite{Kulagin}.   
\beq 
\phi_S (k)&=& g_S\frac{m^2 - k^2}{(\Lambda_S^2 - k^2)^2}  \\ 
\phi_V (k)&=& g_V \frac{m^2 - k^2}{(\Lambda_V^2 - k^2)^{3.5}}  
\eeq 
Here coupling constants $g_S, g_V$ are determined by the normalization of the  
unpolarized distribution $f_1(x)$.   
These functions ensure the correct behavior of the quark distributions at  
large $x$, $f_1^u(x) \sim (1-x)^3$ and $f_1^d(x) \sim (1-x)^4$, which are 
consistent with the standard parametrization\cite{CTEQ}.   On the other hand, 
such large-$x$ behavior does not agree with the parametrization of 
Brodsky {\etal}\cite{Brodsky}.  Recent reanalysis of the experimental
data by Melnitchouk and Thomas\cite{Mel_Thom}, who take into account the Fermi
motion, nuclear binding and nucleon off-shell effects in the deuteron, 
is consistent with the QCD inspired fit by Brodsky {\etal}\cite{Brodsky}.

Taking into account above conditions, we determine the  
model parameters by the experimental data.  We take masses of the  
constituents,  
$m=450\MeV$, $m_S = 700 \MeV$, and $m_V=800 \MeV$ for the quark and  
diquarks.  Values of the cutoff for the quark-diquark vertices are  
$\Lambda_S= 0.49 \GeV$ and $\Lambda_V = 0.66 \GeV$.

The resulting momentum fraction carried by the $u$ and $d$ quarks  
is found to be  0.57 and 0.24 respectively, with the remaining $20\%$  
being left for the gluons and sea quarks.  
The spin fractions of the nucleon become    
$\Delta u = 0.95$ $\Delta d = -0.30$, which yield the nucleon axial-vector
coupling $g_A = 1.25$.   
Calculated total nucleon spin is $\Sigma = 0.65$, which is similar with 
relativistic quark model calculations and larger than the empirical value.   
We also get the tensor charge, $\delta u = 1.17$  and $\delta d =-0.26$.    
We have a relation $\Delta u < \delta u$ independent of the choice of the 
parameters.  The $d$-quark case is more subtle as mentioned above.   
We get $|\delta d| < | \Delta d |$, which seems to be consistent with  
the recent lattice result\cite{h1lattice}  and  
disagrees with the bag model calculation\cite{JaffeJi}.

Such a tendency is quite different from the bag model result, which gives the  
universal inequality $\delta q  > \Delta q$ (\ref{bag2}).    
In the relativistic quark potential model such as the  
bag model, quarks move 
inside the potential independently, and hence there is no 
correlation between the spin of the struck quark and  
spin structure of  remaining spectator 2-quark, except for the trivial  
spin-isospin factor of the nucleon wave function.   
On the other hand, in a bound state of the quark and spin-1 diquark of the 
nucleon, helicities of the struck quark and spectator spin-1 diquark are  
correlated depending on the dynamics of the vertex function.   
Hence, it is rather reasonable that magnitudes of $h_1^d(x)$ and $g_1^d(x)$  
are parameter dependent.

\vspace{1cm} 
 
\ni 
{\bf 3 Momentum evolution and comparison with experiments}

We have obtained the quark distribution functions in the nucleon at the low  
energy model scale $\mu$.   
In order to compare then with experimental data, we carry out  
the $Q^2$ evolution of the distribution function with help of the  
perturbative QCD.  It can be done by using the Altarelli-Parisi  
equation for $f_1(x)$ and $g_1(x)$\cite{AP}.   
As for the transversity distribution,  
the Altarelli-Parisi splitting kernel is developed by Artru and 
Mekhfi\cite{Artru}.  
\beq 
{d h_1(x,Q^2)  \over {d(\ln Q^2)}}={{\alpha _s(Q^2)}  
\over {2\pi }}\int\limits_x^1 {{{dy} \over y}} 
\;  P_T \left( {{x \over y}} \right)\;h_1(y,Q^2) 
\eeq 
The splitting function $P_T$ is given by, 
\beq 
P_T (z)={4 \over 3}\left[ {{2 \over {(1-z)_+}} 
-2+{3 \over 2}\delta (z-1)} \right] 
\label{h1_kern} 
\eeq 
which differs from the $f_1(x)$ and $g_1(x)$ cases.  
Due to the current conservation, the 1st moments of $f_1(x)$ and $g_1(x)$ are 
unchanged by the QCD evolution.  
However, the 1st moment of $h_1(x)$, 
tensor charge, decreases with increasing the momentum scale.

We take the low energy model scale $\mu^2 = 0.2 \GeV^2$ with the QCD cutoff  
$\Lambda = 0.25 \GeV$, which are adopted in Ref.~\cite{Gluck}, and use  
the leading order (LO) evolution equations.   
As discussed in section 2, the low energy effective quark model should provide 
quark distributions at a typical hadronic scale $\sim 1 \GeV$.  
However, if we adopted $\mu = 1 \GeV$ as the low energy scale, the resulting
distribution function would disagree with the experiment.  
In order to reproduce the data, the low energy scale must 
be smaller than 1GeV, typically $\simeq  0.5 \GeV$.  It is difficult to 
understand why such a discrepancy arises at the moment.  
Thus, we treat the starting scale $\mu$ as a free parameter so as to reproduce
the experiments.  
Inclusion of the non-trivial structure of the 
constituent quarks might be crucial to study the origin of this 
discrepancy\cite{Kulagin}, which will be discussed below.

Use of the perturbative QCD below 1GeV seems to be questionable.   
However,  inclusion of next-to-leading order (NLO) corrections modifies  
about $10 \sim 20  
\%$ of the LO evolution result, which will be briefly discussed later.   
Of course, if we take much lower scale, {\eg} $\sim 0.1 \GeV^2$,  
then difference of 
the LO and NLO results becomes very large, and the perturbative evolution 
is no longer reliable.   
All the results presented below are evolved to $Q^2= 10 \GeV^2$.    
When we evolve the tensor charge from $\mu ^2 = 0.2 \GeV^2 $ to a few 
$\GeV^2$ scale, the reduction of the tensor charge is less than 20$\%$.

We first show  $f_1^u(x)$ and $f_1^d(x)$ with the CTEQ experimental 
parametrization\cite{CTEQ} in  Fig.3    
at $Q^2 = 10 \GeV^2$ by thick-dashed and thin-dashed curves.     
The resulting distribution at $x \sim 0.3$ is about $40\%$ larger than the  
experimental fit.   In the framework of the quark-diquark model (and most  
effective models), it is quite difficult to get better agreements within the 
reasonable parameters of the model.   
In the next section, we will introduce  
dressing of the constituent quark due to the Goldstone boson  
cloud, which greatly improves the numerical results.

\begin{figure} 
\begin{center} 
\psfig{file=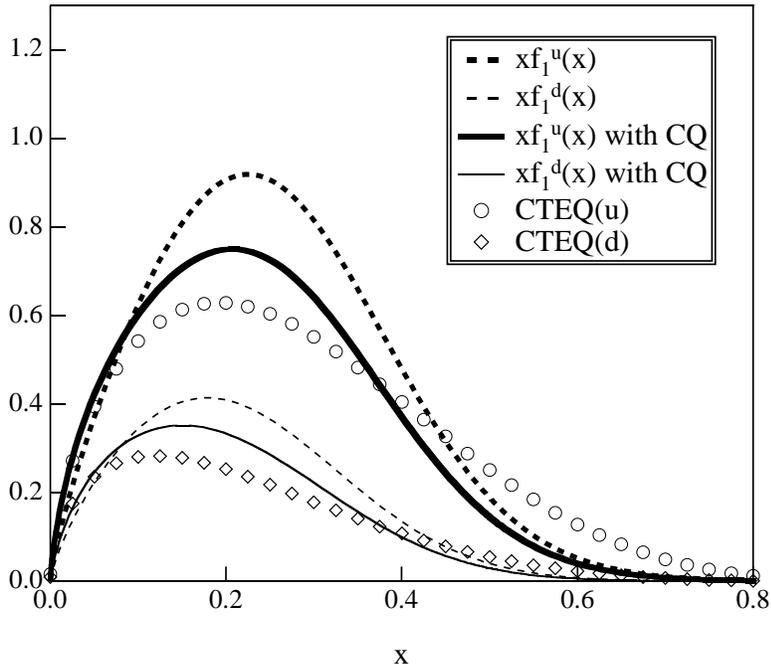,height=3.5in} 
\caption{The unpolarized valence quark distribution of the 
nucleon\label{fig_f1}} 
\end{center} 
Unpolarized valence $u$ and $d$ quark distributions at $Q^2= 10\GeV^2$.    
Thick-dashed and thin-dashed curves indicate results with bare constituent 
quarks for $u$ and $d$ distributions, respectively.   
Results with the GS boson dressing are shown by thick-solid and thin-solid 
curves for $u$ and $d$ quarks.   
The circle denotes the CTEQ4 parametrization for the valence $u$-quark 
distribution, and the boxes for the $d$-quark\cite{CTEQ}.   
\end{figure} 

Here, we comment on the work by Kulagin {\etal} \cite{Kulagin}.  Procedure 
to obtain the unpolarized structure function $f_1(x)$ presented here is  
the same as one of Ref.~\cite{Kulagin}.   
They also estimate contribution from structure of the 
constituent quark itself, which is described by the pion dressing at the 
middle-$x$ region and the Regge exchange contribution at the small-$x$.   
Combining those important contributions with the bare quark-diquark model,  
they have obtained improved quark distributions as the input of the $Q^2$ 
evolution at typical hadronic scale $\sim 1\GeV$, at which the use of the  
perturbative QCD evolution is reasonably acceptable.   
They have reproduced observed structure function of the nucleon very well.

We have analyzed our results for 
the spin independent as well as the spin dependent structure function with  
the same parameter set used in Ref.~\cite{Kulagin}.   
Unfortunately, we find resulting spin dependent structure functions become 
too  small compared with the experimental data.   
This is because they adopt large diquark masses $m_D > 1 \GeV$.   
To reproduce the nucleon axial-vector coupling $g_A$, it is necessary to take 
smaller values of diquark masses around $700\MeV$ within this quark-diquark 
model, and thus impressive agreements of $f_1(x)$ with the data are lost.   
Calculated distributions with our parameters also require the evolution from  
$\mu^2 \sim 0.2 \GeV^2$ to get reasonable agreements with the data.

The longitudinal spin structure function of the proton $G_1^p (x) =  
\frac{1}{2} \left[ \frac{4}{9} g_1^u(x)+ \frac{1}{9} g_1^d(x) \right]$  
and the neutron  
$G_1^n (x) = \frac{1}{2} \left[ \frac{1}{9} g_1^u(x)+ \frac{4}{9} g_1^d(x)  
\right]$ are shown in Fig.4.     
Our calculation reproduces the nucleon axial-vector coupling  
$g_A=1.25$, but the quark spin fraction $\Delta u = 0.95$ and $\Delta d =  
-0.30$  
are  somewhat different from the empirical values $\Delta u = 0.82 \pm 0.03$  
and $\Delta d = -0.43 \pm 0.03$ \cite{Ellis}.    
As a result, calculation overestimates  $g_1^p(x)$ and the total nucleon spin 
fraction $\Sigma = 0.65$ compared with the empirical one $\Sigma = 0.31  
\pm 0.07$.   
Note that  $g_1^d(x)$ becomes positive at the large $x$ in this model, 
which is consistent with the QCD inspired parametrization of  
the helicity distribution functions by Brodsky {\etal}\cite{Brodsky}.   
 
\begin{figure} 
\begin{center} 
\psfig{file=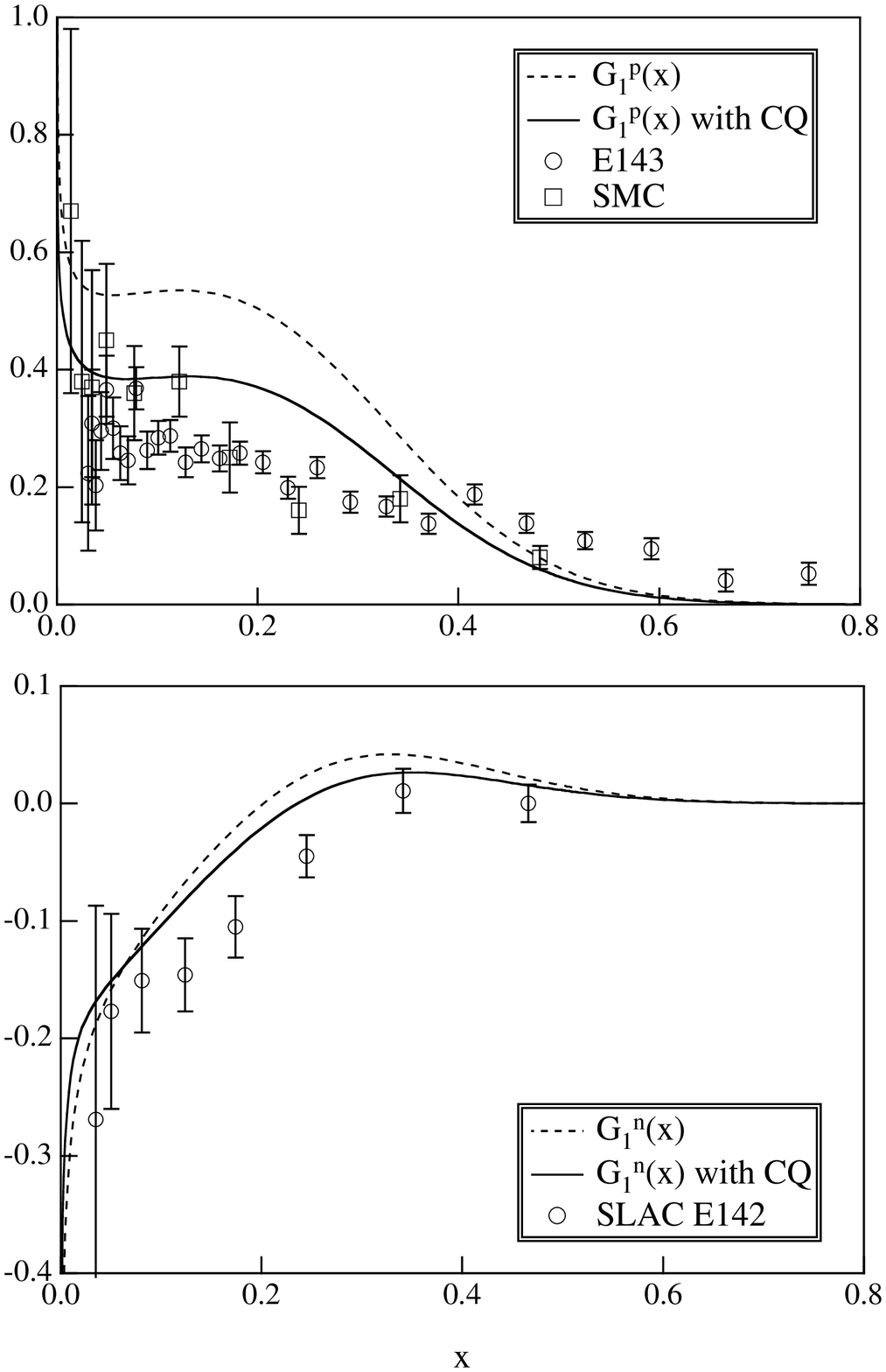,height=6.5in} 
\caption{The longitudinal spin structure function of the proton and the  
neutron\label{fig_g1}} 
\end{center} 
\vspace{-0.2cm} 
Upper figure shows the proton spin structure function  $G_1^p(x)$,  
and the lower one the neutron $G_1^n(x)$ at $Q^2 = 10 \GeV^2$.   In each
figure, result with the bare quark is shown by the dashed curves, and one
with the GS boson cloud by the solid curves.   
Experimental data is taken from Ref.~\cite{Exp_g1}.   
\end{figure} 

Before discussing the chiral-odd structure function, we point out the  
importance of ratio of the structure functions in order to extract the 
non-perturbative aspects of QCD from the deep inelastic data.    
As mentioned above,  
shapes of calculated quark distribution depend so strongly on the choice of 
the low energy scale that it is not easy to draw definite conclusions from  
the comparison of the model calculations with experiments.   
However, if we take a ratio of several distribution functions,   
the calculated ratio itself is rather insensitive to the choice of the  
scale $\mu ^2$, since  
ambiguities from the evolution are considerably canceled out by taking a 
ratio.   
Therefore, we can know how non-perturbative structure of the nucleon affects 
the distribution functions in the deep inelastic scattering.   
Also, by taking a ratio, we can check internal consistency of the model 
calculations.   
Results for $f_1^d(x) / f_1^u(x)$ and $A_1^p =  G_1^p(x) /F_1^p(x)$  
presented in Fig.5   
show remarkable agreements with the data.  
These agreements are consequence of the spin-flavor structure of the 
nucleon, which is incorporated by the quark-diquark model\cite{MM}.   
If we took un-physical values for diquark masses, {\eg}  
$m_S \ge m_V$,  agreement would be lost.  Correct spin-flavor 
correlation of the 
nucleon wave function leads to a good description of these ratios.

\begin{figure} 
\begin{center} 
\psfig{file=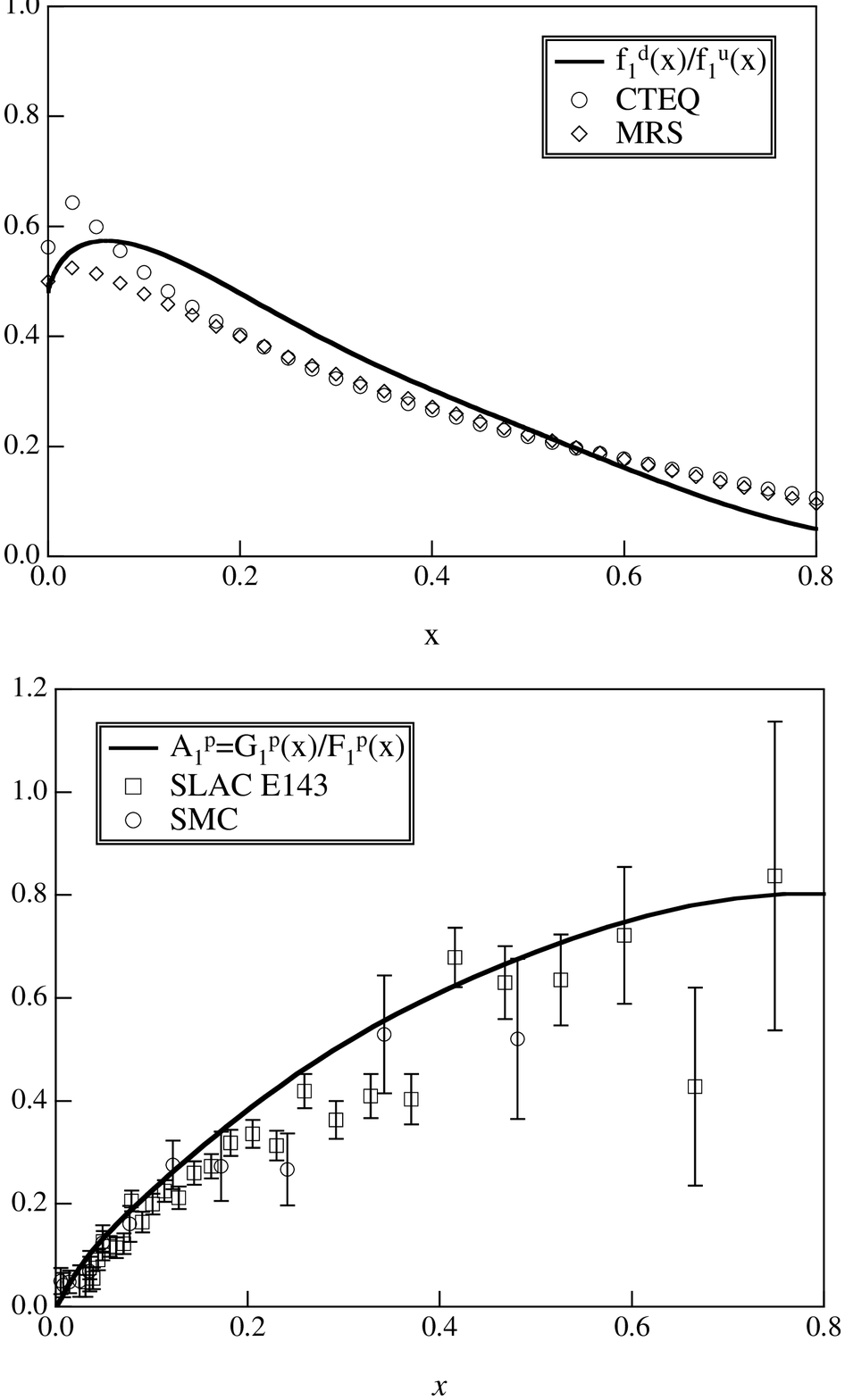,height=6.5in} 
\caption{Ratios of the structure functions with the experimental  
data\label{fig_ratio}} 
\end{center} 
\vspace{-0.2cm} 
$f_1^d(x) / f_1^u(x)$ and $G_1^p(x)/F_1^p(x)$ are shown in upper and  
lower figures, respectively.  In each figure, calculated result is shown by 
the solid curve with the data\cite{CTEQ,Exp_g1}.    
\end{figure}

We present $h_1^u(x)$ and $h_1^d(x)$ in Fig.6   
at $Q^2 = 10 \GeV^2$.    
The $u$-quark distribution shows similar behavior with $f_1(x)$ and $g_1(x)$,  
and is satisfied with $f_1^u (x) > h_1^u(x) > g_1^u (x)$.     
On the other hand,   shape of the $h_1^d(x)$ distribution differs from that of 
$g_1^d(x)$, and their magnitudes are similar.  
We will discuss the flavor dependence of the transversity spin distribution 
function in detail after introducing the GS boson dressing in the next 
section.  

\begin{figure} 
\begin{center} 
\psfig{file=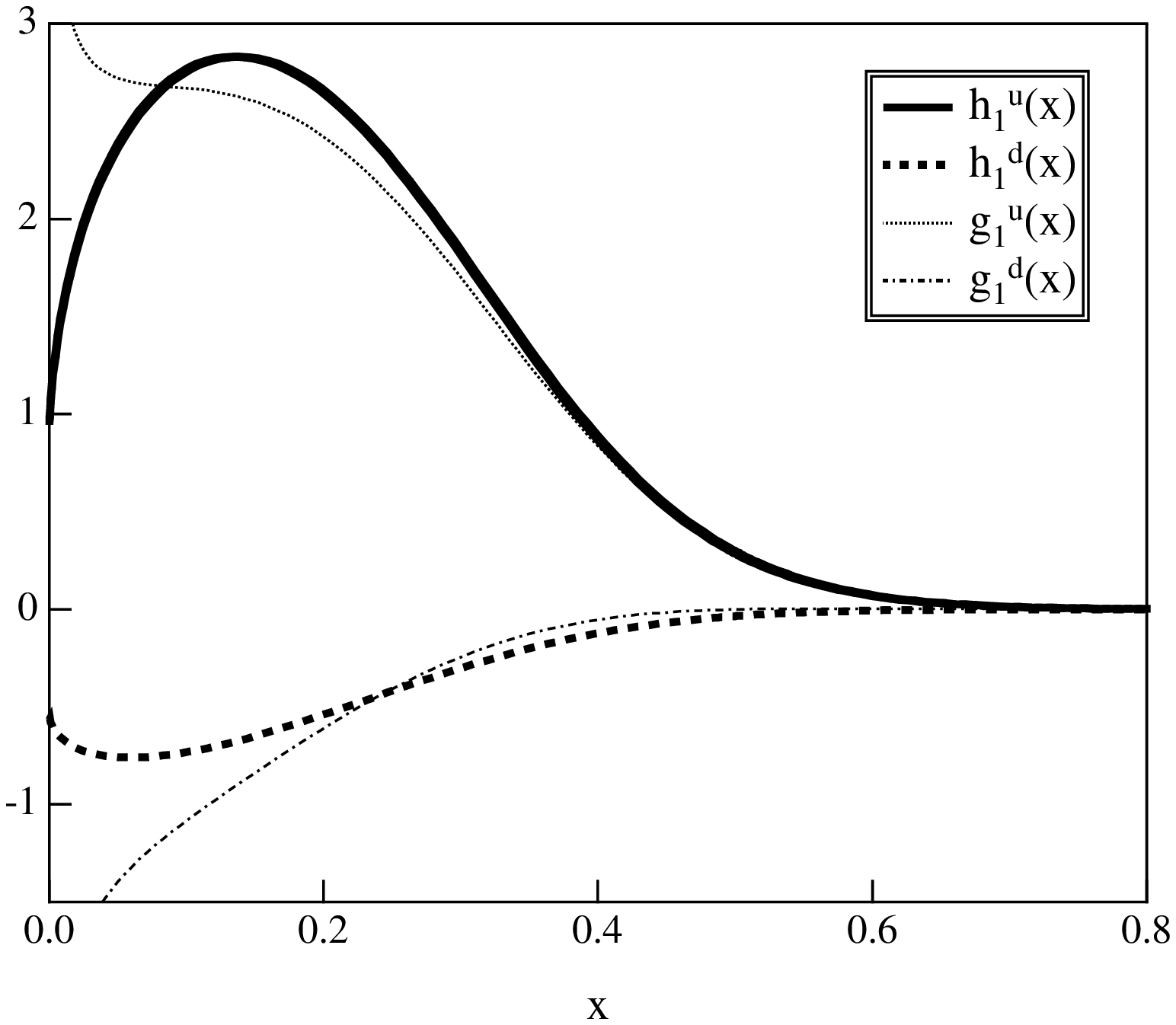,height=4in} 
\caption{The transversity spin structure function $h_1(x)$ for the $u$ and  
$d$ quarks\label{fig_h1}} 
\end{center} 
\vspace{-0.2cm}
$h_1^u(x)$ and $h_1^d(x)$ are displayed by the thick solid  
and dashed curves, 
respectively.  For comparison, $g_1^u(x)$ and $g_1^d(x)$ are shown by the  
dotted and dash-dotted curves.   
\end{figure} 

We remark that the helicity distribution functions $g_1(x)$ becomes much 
larger 
than the transversity distributions $h_1(x)$ at the small-$x$ region, $x<0.1$. 
Remember that, at the low energy scale, $g_1^u(x)$ is smaller than  
$h_1^u(x)$ for all $x$, as shown in Fig.2.    
This smallness of $h_1(x,10\GeV^2)$ at $x<0.1$ is due to the difference 
of the QCD evolution.     
Regarding the 1st moment of $h_1(x)$ structure function, the splitting kernel  
Eq.~(\ref{h1_kern}) for $h_1(x)$ does not give a significant difference,   
{\ie} the reduction of the 1st moment is less than  20$\%$ by carrying 
out the $Q^2$ evolution from $\mu^2$ to $10\GeV^2$.   
At small-$x$ region, however, the $Q^2$ evolution procedure generates  
a quite large difference between $g_1(x)$ and $h_1(x)$.   
As discussed in Ref.~\cite{pp}, such a difference produce great influence 
on the 
$pp$ Drell-Yan experiment, which is planed at RHIC-SPIN.      
Transverse Drell-Yan spin asymmetry $A_{TT}$ becomes substantially smaller 
than the longitudinal 
one $A_{LL}$, which makes it very hard to measure the transversity spin 
distribution function $h_1(x)$ at RHIC.

\vspace{1cm} 
 
\ni 
{\bf 4 Corrections from the constituent quark structure}

In the previous section, we treat the constituent quark as a point-like
particle.  
Here, we shall consider contributions from the Goldstone boson dressing of the 
constituent quark studied in Ref.~\cite{Suzuki}.   
In the chiral quark model of Manohar and Georgi\cite{CQM},  
the constituent quark interacts 
with the Goldstone (GS) bosons, $\pi$, $K$ and $\eta$,  associated with the  
spontaneous chiral symmetry 
breaking.  Within the one GS boson approximation, diagrams illustrated in 
Fig.7 contribute to the structure function.    
The GS boson fluctuation around 
the constituent quark changes probability to find a bare quark state,  
and gives 
soft contribution to the quark distribution function.  Also, it changes the 
CQ spin structure by emitting the GS boson into the $P$-wave state.

\begin{figure} 
\begin{center} 
\psfig{file=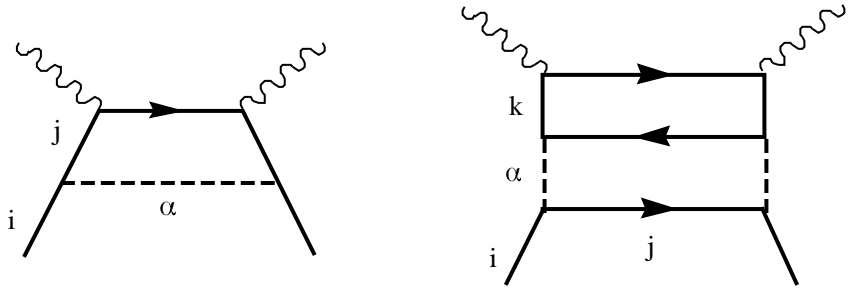,height=1.5in} 
\caption{Goldstone boson corrections to the quark distribution function 
\label{fig_cqm}} 
\end{center} 
\vspace{-0.2cm}
Fig.\ref{fig_cqm}(a) indicates the Goldstone boson spectator  
process, and  
Fig.\ref{fig_cqm}(b) probes the internal structure of the Goldstone boson  
with the constituent quark being spectator.    
The constituent quark and the Goldstone boson are depicted by the solid and 
dashed lines, respectively.   
\end{figure} 

The constituent quark-GS boson splitting functions giving corrections to the 
twist-2 structure functions are given by\cite{Suzuki},  
\beq 
P(y)_{j \, \alpha / i}={g^2 \over {8\pi ^2}}   
\int_{}^{} {dk_T^2}{1 \over {y(1-y)}} 
{{(m_j - m_i y)^2 + k_T^2} \over {y\left[ {m_i^2-M^2_{j+\alpha}} \right]^2}} 
\label{spil-f} 
\eeq 
\beq 
\Delta P(y)_{j \, \alpha / i}={g^2 \over {8\pi ^2}}  
\int {dk_T^2}{1 \over {y(1-y)}} 
{{(m_j - m_i y)^2 - k^2_T } \over {y\left[ {m_i^2 - M^2_{j+\alpha}} 
\right]^2}} 
\label{spil-g} 
\eeq 
\beq 
\delta P(y)_{j \, \alpha / i}={g^2 \over {8\pi ^2}}  
\int_{}^{} {dk_T^2}{1 \over {y(1-y)}} 
{{(m_j - m_i y)^2 } \over {y\left[ {m_i^2-M^2_{j+\alpha}} \right]^2}} 
\label{spil-h} 
\eeq 
for $f_1(x)$, $g_1(x)$ and $h_1(x)$ distribution functions,  
respectively.   
Here, $m_i, m_j, m_{\alpha}$ are masses of the $i,j$- constituent  
quarks and  
the pseudoscalar meson $\alpha$, respectively.   
$M^2_{j \, \alpha } = \frac{m_j^2+k_T^2}{y}  
+ \frac{m_{\alpha}^2 + k^2_T}{1-y}$ is the invariant mass squared of the final 
state.   
We use the quark-GS boson coupling constant $g = 3.76$ as 
suggested by the effective theories\cite{NJL}.   
These functions are calculated in the Infinite Momentum 
Frame,\footnote{It is possible to calculate the dressing in the covariant
approach, but convolution breaking terms appear and the resulting formulae  
become 
much complicated.  It is also possible to evaluate both bare distributions and
dressing in the infinite momentum frame from the beginning.} 
exponential cutoff is used with  
$\Lambda_{GS} = 1.9 \GeV$, which is determined to reproduce the  
violation of the Gottfried sum rule\cite{Suzuki}:    
\beq
g \ra g \;\, \mbox{exp} \left[ \frac{m_i^2 -M^2_{j \, \alpha }} 
{2 \Lambda_{GS} ^2} \right]  \; .
\label{cutoff-f}
\eeq

In terms of these splitting functions,  
the quark distribution $q_j(x)$ generated 
by the emission of the GS boson $\alpha$ from a parent quark distribution $q_i 
(x)$ (Fig.7(a)) is expressed as;  
\beq 
q_j (x) = 
\int_x^1 {{dy} \over y} \,P_{j \, \alpha / i} (y) \, q_i \left( {{x \over y}}  
\right) 
\eeq 

The CQ spectator process in Fig.7(b), which contributes to 
the small $x$ region, is given by,  
\beq 
q_k (x) = 
\int {{{dy_1} \over y_1}}  \, {{{dy_2} \over y_2}} 
\, V_{ k / \alpha} \left( \frac{x}{y_1} \right) P_{\alpha \, j / i}  
\left( \frac{y_1}{y_2} \right) \,q_i (y_2)  
\eeq 
where $V_{q/ \alpha}(x)$ is the quark distribution function in the GS 
bosons.   
The splitting function is obtained by a symmetric relation as  
$P_{\alpha \, j / i} (x) = P_{ j \, \alpha / i} (1-x)$.     
Since we incorporate only the pseudoscalar mesons, this 
diagram never contributes to the spin-dependent processes.

Combining the flavor structure of the GS bosons,  
one can rewrite the twist-2 structure functions renormalized by the GS boson 
clouds as follows, 
\beq 
f_1^u (x)_R &=& Z f_1^u (x)+ P_{u \pi / d} \otimes f_1^d +  
V_{u / \pi }\otimes P_{\pi \, d / u} \otimes f_1^u+ 
{1 \over 2} P_{u \, \pi/ u} \otimes f_1^u  \neqn \\ 
 & &+ {1 \over 4}V_{u / \pi }\otimes P_\pi \otimes (f_1^u+f_1^d)+  
V_{u / K}\otimes P_{K \, u / u }\otimes f_1^u \neqn \\ 
& & + \frac{1}{6} P_{u \, \eta/ u} \otimes f_1^u +  
 {1 \over 36}
V_{u / \eta }\otimes P_\eta \otimes (f_1^u + f_1^d) 
\label{f1r} 
\eeq 
\beq 
f_1^{\bar u} (x)_R &=& V_{u / \pi} \otimes P_\pi \otimes f_1^d  \neqn \\ 
&& \hspace{0.6cm}+ \frac{1}{4} V_{u / \pi} \otimes P_\pi \otimes (f_1^u+ 
f_1^d) + \frac{1}{36} V_{u / \eta} \otimes P_\eta \otimes (f_1^u + f_1^d) 
\label{f1rbar} 
\eeq
Similar expressions can be written for the $d$-quark case.  
For the spin-dependent distribution functions,  
\beq 
g_1^u (x)_R = Z g_1^u (x) + P_{u \pi / d} \otimes g_1^d 
+{1 \over 2}P_{u \, \pi/ u} \otimes g_1^u  +  
\frac{1}{6} P_{u \, \eta/ u} \otimes g_1^u 
\label{g1ur} 
\eeq 
\beq 
g_1^d (x)_R =  Z g_1^d (x) + P_{d \pi / u} \otimes g_1^u  
+{1 \over 2}P_{d \, \pi/ d} \otimes g_1^d  
+ \frac{1}{6} P_{d \, \eta/ d} \otimes g_1^d 
\label{g1dr} 
\eeq 
Here, $\otimes$ expresses the convolution integral.  
Formulae for $h_1(x)$ distribution are obtained by just replacing 
$g$ with $h$ in Eqs.~(\ref{g1ur}),(\ref{g1dr}).  
The renormalization constant $Z$ is given by, 
\beqn 
Z = 1 - \frac{3}{2} \bra P_\pi \ket - \bra P_K \ket - \frac{1}{6} \bra P_\eta 
\ket  
\eeqn 
where $\bra P_\alpha \ket$ is the 1st moment of the unpolarized 
splitting function.   Within our model, $Z$ is found to be 0.76.    
We emphasize that $\Delta P$ is somewhat negative or nearly zero, while  
$\delta P$ is positive \cite{Suzuki}.   
Calculated first moments of the splitting functions for  
pion case are $\bra P_\pi \ket = 0.13$, $\bra \Delta P_\pi \ket = -0.06$  
and $\bra \delta P_\pi \ket = 0.07$.  This sign difference of the helicity and  
transversity distributions causes considerable effects on the  
proceeding calculations.

The results for the unpolarized valence quark distribution defined  
by $f_1^q(x) - f_1^{\bar q} (x)$ are presented in Fig.3 by using  
Eqs.~(\ref{f1r}, \ref{f1rbar}).   
Here, we show the original result by the dashed curves, and the  
results with 
the CQ structure by the solid curves.  The QCD evolution procedure is the same 
as before with $\mu ^ 2 = 0.2 \GeV^2$.   
The renormalization of the bare quark distribution brings down the peak of the 
distribution function for both $u$ and $d$ quarks, and  
the agreement around $x \sim 0.1$ is considerably improved due to the GS 
boson dressing.

The GS boson contribution to $g_1^u$ and $g_1^d$ are different, as shown in 
Fig.8.   
As a result of the renormalization,  $g_1^u(x)$ distribution function 
decreases substantially, while $g_1^d(x)$ is almost unchanged.   
This fact is easily understood as follows.   
The renormalization factor $Z$ always reduce absolute values of the 
distribution function, {\ie}~positive contribution to original (negative)  
$g_1^d(x)$.  The largest correction to $g_1^d(x)_R$,   
the second term of Eq.~(\ref{g1dr}) 
$\Delta P_{d \pi / u} \otimes g_1^u$, gives a negative contribution, since  
$\Delta P_{d \pi / u}$ is negative.   
Thus, cancelation of the renormalization effect and the correction term 
keeps $g_1^d(x)_R$ almost unchanged.   
Resulting spin fractions are tabulated in Table 1 with the original 
values and  
lattice data.

\begin{figure} 
\begin{center} 
\psfig{file=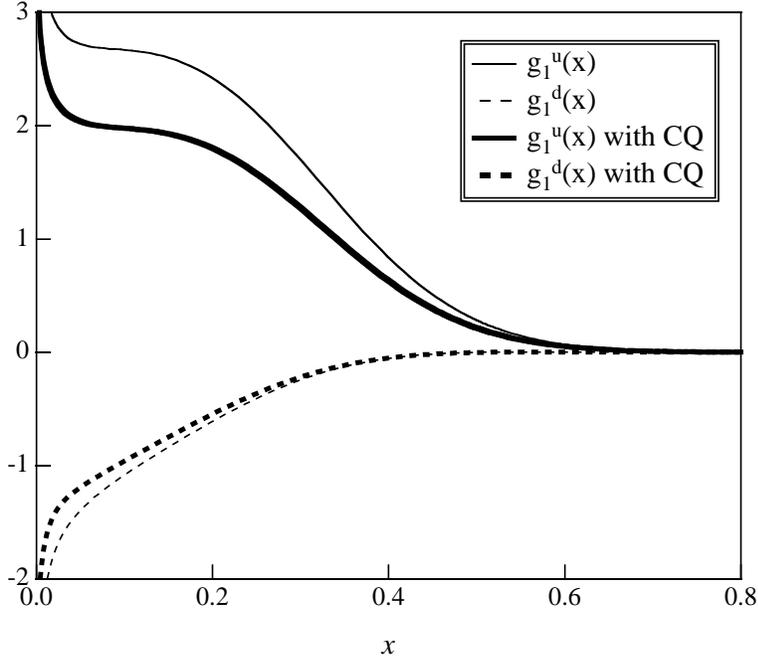,height=3.5in} 
\caption{Helicity distribution functions $g_1^u(x)$ and 
$g_1^d(x)$ with the GS boson corrections\label{fig_g1c}}
\end{center}
\vspace{-0.3cm}   
Results with the dressing are shown 
by the thick solid and dashed curves for $u$ and $d$ quarks, and ones  
without the dressing by thin curves.
\end{figure}

This is not the case for the transversity spin distribution.   
Results are shown in Fig.9 with the original ones  
and $g_1^d(x)_R$ for comparison.   
Both $h_1^u(x)_R$ and $h_1^d(x)_R$ decrease, particularly, the $d$-quark 
transversity 
spin distribution becomes about a half of the result without the GS boson 
dressing.   
This is because the GS-boson splitting function $\delta P$ is positive for the 
transversity spin case, opposite to the helicity distribution case.   
Hence, for the $d$-quark, the renormalization as well 
as a correction term $\delta P_{d \pi / u} \otimes h_1^u$ are positive,  
which drive 
further reduction of the $d$-quark transversity spin distribution.   
At $x \sim 0.1$, the transversity spin distribution becomes the one-third of 
the helicity distribution.  
Corrected tensor charges are given in Table 1.

\begin{figure} 
\begin{center} 
\psfig{file=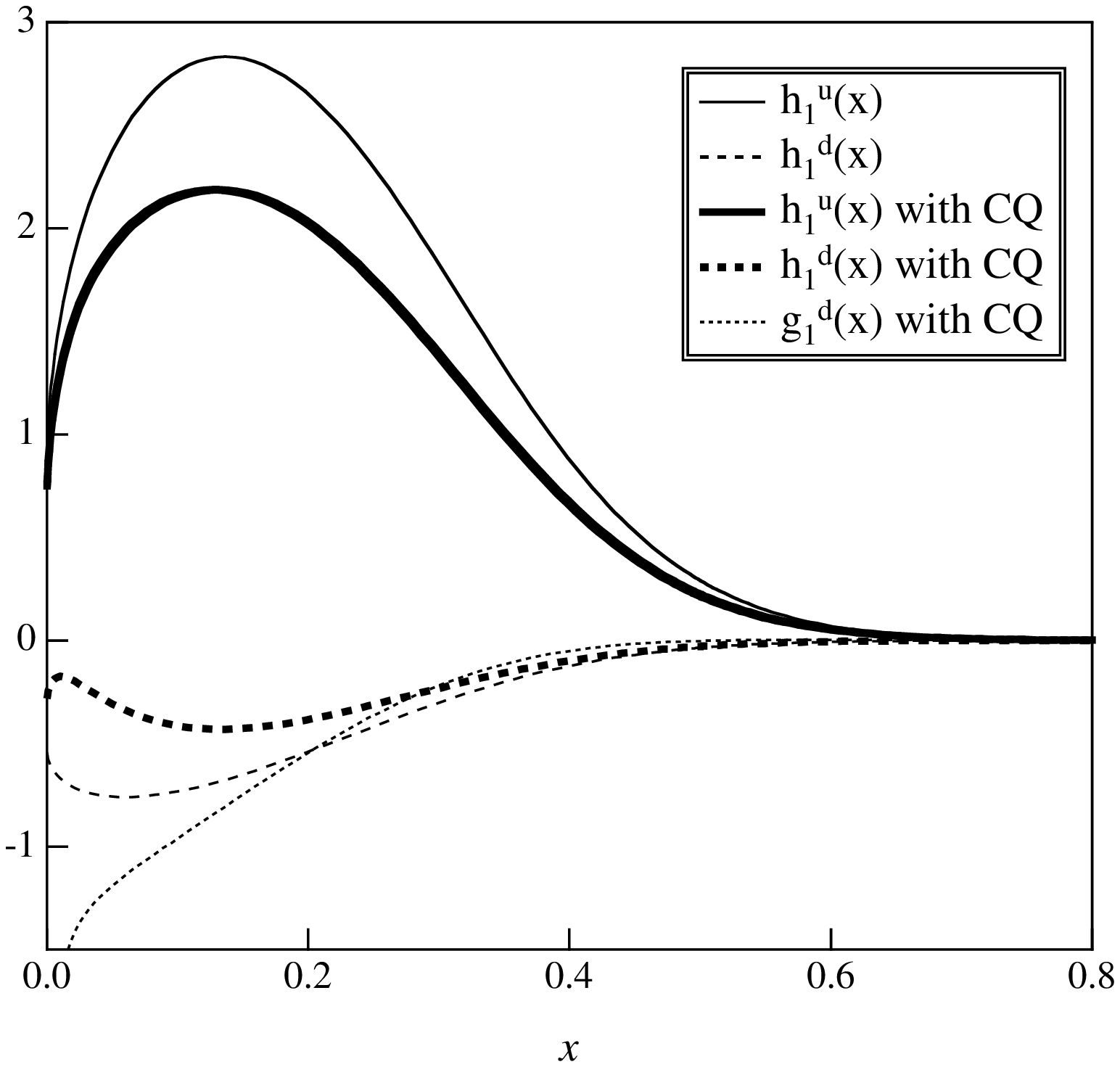,height=4.5in} 
\caption{Transversity spin distributions $h_1^u(x)$ and $h_1^d(x)$ 
with the GS bosons\label{fig_h1c}} 
\end{center} 
\vspace{-0.2cm}
Notations are same as those in Fig.\ref{fig_g1c}.   
We also show  $g_1^d(x)$ with the GS boson dressing by the dotted 
curve.   
\end{figure} 

It is very interesting and important to consider 
experimental possibilities to measure such a flavor dependence of the 
transversity spin structure function.  
The Drell-Yan process, which is planned in RHIC-SPIN or 
HERA-N is not suitable to determine the flavor decomposition of the 
structure function, since all possible combination of $q$-$\bar q$ pair
contributes to the $\mu^+ \mu^-$ cross section.  
On the other hand, semi-inclusive deep inelastic scattering may be 
a good candidate to observe the flavor dependence of the spin structure
function.  
For example, according to the work of Kotzinian and Mulders \cite{Kotz}, 
we stress the
semi-inclusive deep inelastic scattering on the transversely polarized nucleon.
Let us consider the $\pi$ production process (momentum $P_\pi$)  on the proton 
target (momentum $P$) with unpolarized lepton beam (momentum $l$).   
They have found the following weighted transverse
asymmetry of the cross section is related with the transversity spin
distribution of the nucleon.  
\beq
A_T(x,y,z,S_T) = \frac{ \int d \phi_l \int d^2 P_{\pi \bot} 
\frac{| P_{\pi \bot} |}{z M_\pi} \sin (\phi_s^l + \phi_h^l) (d \sigma ^+ 
-d \sigma ^-)} {\int d \phi_l \int d^2 P_{\pi \bot}(d \sigma ^+ + d \sigma ^-)}
\label{asym}
\eeq
where scaling variables are given by $x=Q^2 / 2 P \cdot q$, $y=P \cdot q /
P \cdot l$, and the momentum fraction of a produced pion 
$z = P \cdot P_\pi / P \cdot q$.  $P_{\pi \bot}$ is the transverse momentum
perpendicular to $\bf q$, and azmitual angles $\phi_s^l$, $\phi_h^l$  are 
defined with respect to the lepton plane.  
Asymmetry of this cross section is generated by the so called Collins effect in
the fragmentation process\cite{Collins}.   
Eq.~(\ref{asym}) is expressed with the quark distribution and 
fragmentation functions as,
\beq
A_T(x,y,x) = -|S_T | \frac{2 (1-y) }{1+(1-y)^2} 
\frac{h_1^a(x) H_{1 \, \bot} ^a (z)} {f_1^a(x)  D_1^a (z)}
\eeq
where $h_1^a(x)$ and $f_1^a(x)$ are the quark distribution functions, and 
$D_1^a (z)$ and $H_{1 \, \bot} ^a (z)$ unpolarized and transversely polarized
quark fragmentation functions from a quark flavor $a$ to the pion.  
It is well known that the $\pi^+$ production process is dominated by the 
$u$-quark fragmentation, and $\pi^-$ by the $d$-quark.  
Also, since pions are isospin symmetric objects, the quark fragmentation 
functions are the same for $u \ra \pi^+$ and $d \ra \pi^-$.  
Therefore, if we take a ratio of asymmetries for $\pi^+$ and $\pi^-$ production
processes under above assumptions, we can arrive at the following expression.
\beq
\frac{A_T^{\pi^-}}{A_T^{\pi^+}} (x)
 = \frac {h_1^d(x)}{h_1^u(x)} \frac{f_1^u(x)}{f_1^d(x)}
\eeq
where unknown `transversely polarized quark fragmentation function' is canceled
out.  
$f_1^u(x) / f_1^d(x) $ is well determined so far, thus we can extract a 
ratio of the transversity spin distribution functions of $u$ and $d$ quarks
by counting the numbers of produced $\pi^+$ and $\pi^-$.    
Using a similar technique with 
the polarized lepton beams on the longitudinally polarized proton, 
it may be possible to obtain a ratio of helicity distribution 
functions \cite{Kotz}.

We show in Fig.10 $h_1^d(x)/ h_1^u(x)$ using our calculated
transversity distributions, where we can see clear     
flavor dependence of the spin-dependent structure functions.    
We also show  $g_1^d(x) / g_1^u(x)$ in our model, and one obtained by 
the parametrization of Brodsky {\etal}\cite{Brodsky}.  
Due to the reduction of $h_1^d(x)$, $h_1^d(x)/ h_1^u(x)$ is much smaller than  
$g_1^d(x) / g_1^u(x)$ around $x \sim 0.1$.   
Calculated $g_1^d(x) / g_1^u(x)$ is consistent with the 
the parametrization of  Brodsky {\etal}\cite{Brodsky}.   
The bag model calculation gives similar behavior for  
$h_1^d(x)/ h_1^u(x)$ and $g_1^d(x) / g_1^u(x)$.   
Note that, in the naive $SU(4)$ symmetry limit, 
$h_1^d(x) /  h_1^u(x) = g_1^d(x)/  g_1^u(x) = -1/ 4$.  
Alternative experimental processes to measure $h_1(x)$ 
are now under investigation.   

\begin{figure} 
\begin{center} 
\psfig{file=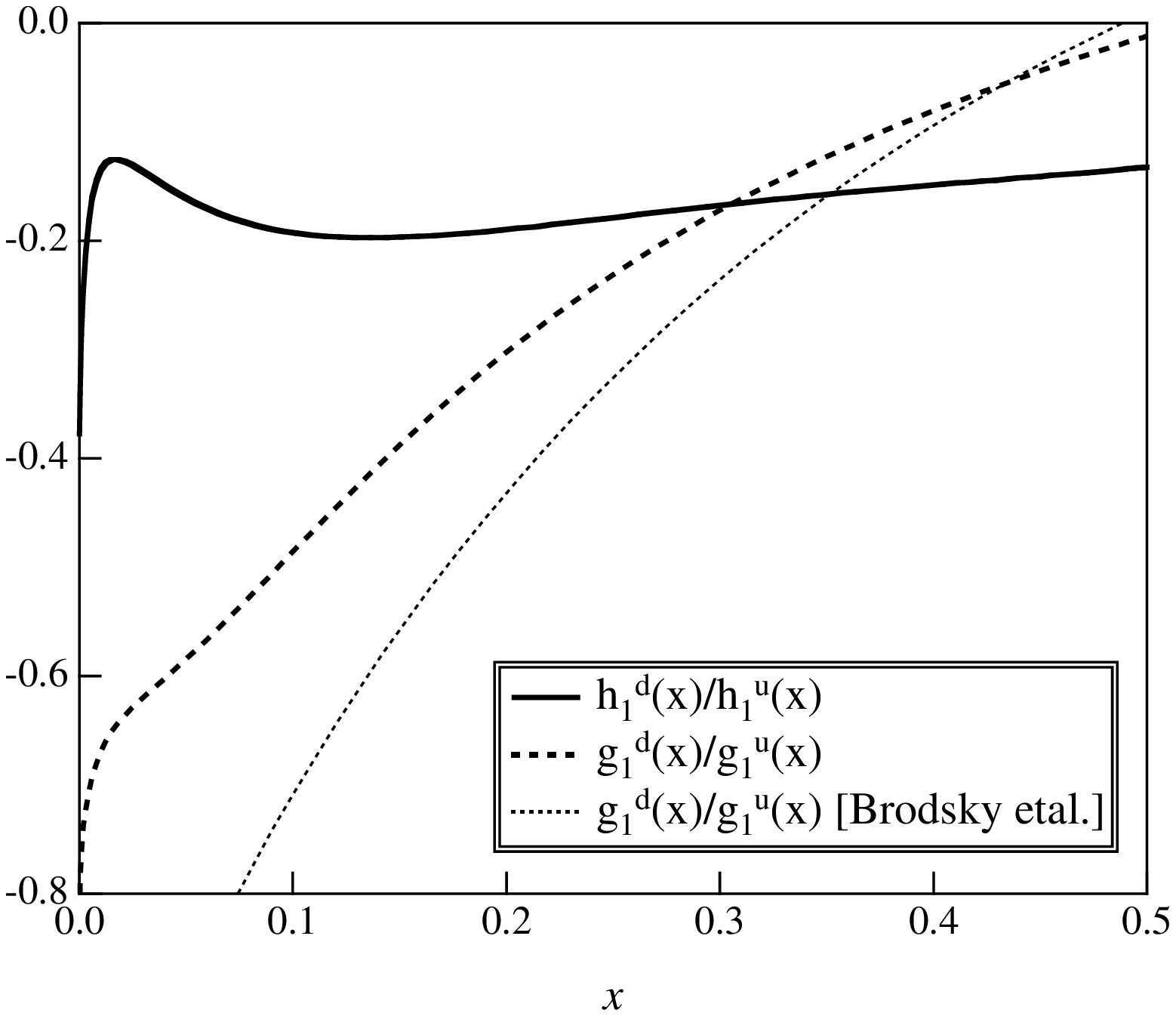,height=4.5in} 
\caption{Flavor dependence of the spin dependent structure  
functions\label{fig_hdhu}} 
\end{center} 
\vspace{-0.2cm}Ratios $h_1^d(x)/h_1^u(x)$ and $g_1^d(x) / g_1^u(x)$ are  
shown by the  
solid and dashed curves.  The QCD inspired parameterization\cite{Brodsky} for  
$g_1^d(x) / g_1^u(x)$ is also shown by the dotted curve.   
\end{figure}

\vspace{1cm}

\ni 
{\bf 5 Discussions}

We have studied the twist-2 quark distribution functions $f_1(x)$, $g_1(x)$  
and $h_1(x)$ of the nucleon in terms of the 
phenomenological quark-diquark model.    We deal with the scalar and 
axial-vector effective nucleon-quark-diquark vertices from the symmetry  
consideration.   
Resulting distribution functions show a saturation of  the Soffer's  
inequality $ f_1(x) + g_1(x) = 2 h_1(x)$ when we use the scalar type of the  
quark-diquark interaction,  
whereas the axial-vector case is not saturated.   
Moreover, the spin distribution function satisfies a universal inequality  
$g_1(x) < h_1(x)$ for the scalar vertex case, which is  
similar with 
result of the relativistic potential model like the bag model.    
On the other hand, relative size of these distribution functions depends 
on the 
model parameters in the axial-vector vertex case.  With the model parameters 
fixed to reproduce the known observables, the quark momentum fraction and  
the nucleon axial-vector coupling $g_A$,  
we have found that $|h_1(x)|$ is smaller 
than $|g_1 (x)|$ or they are comparable at least.    
Since the axial-vector case contributes to only the $d$-quark distribution from 
the $SU(6)$ symmetry, this behavior is consistent with the recent lattice QCD 
calculations\cite{h1lattice} and the QCD sum rule result\cite{QSR},  
where the $d$-quark tensor charge is 
substantially suppressed compared with the axial-charge.

This difference simply comes from the spin structure of the quark-diquark  
vertices.  We have checked other types of the Dirac structure for the vertex,  
and found, if the vertex is independent of the spin of the spectator diquark,  
{\eg} scalar, pseudo-scalar,  derivative coupling, universal 
relations Eqs.~(\ref{bag1},\ref{bag2}) hold.   
However, for the vector or axial-vector interactions,  
such relations are not realized, depending on the dynamics of the 
quark-diquark system.

We have also estimated corrections due to the non-trivial structure of the 
constituent quark itself.  The constituent quark and Goldstone bosons are 
believed to be most important degrees of freedom  
below the chiral symmetry breaking scale $\sim 1\GeV$.   
Corrections from the GS boson fluctuation can not be 
negligible, because it produces a renormalization of the CQ state as 
well as it changes the CQ spin structure by emitting the GS boson into the 
$P$-wave state relative to CQ.    
The probability to 
find a bare constituent quark decreases to about $70 \%$, and the quark 
distribution function becomes soft.   
Although corrections to the spin dependent parts are parameter  
dependent, the GS boson fluctuation gives positive contributions to the 
transversity spin distribution $h_1(x)$, while negative for the helicity 
distribution $g_1(x)$.   
This feature causes further reduction of the $d$-quark tensor charge as shown 
in Table 1.

Recently, experimental methods to test the nucleon transversity spin  
distribution are discussed by several authors\cite{JaffeJi,Collins,exp}.   
The transversely polarized $pp$ Drell-Yan process is planed at RHIC, but 
it is suggested that the transverse double spin asymmetry becomes quite small 
in the accessible kinematic region at RHIC\cite{pp}.  
This is due to the suppression of  $h_1(x)$ at $x < 0.1$ shown in 
Fig.6.  
This is a general result of the effective quark model calculation, if 
one accepts calculations of the effective theory 
as quark distributions at the low energy scale $\mu \sim 0.5 \GeV$.   
The QCD evolution procedure from $\mu^2$ to the experimental scale generates 
a large difference between $h_1(x)$ and $g_1(x)$ at $x<0.1$, 
even if the effective quark model provides the same inputs for $h_1(x)$ 
and $g_1(x)$.

In Ref.~\cite{Single}, Artru {\etal}~have given a constraint for the  
transversity spin distribution in the nucleon from the analysis of the  
single spin asymmetry in 
the pion production in the transversely polarized proton-proton 
collision $p + p \uparrow \ra \pi + X$.   
Their best fit gives 
\beq 
\frac{h_1^u(x)} {f_1^u(x)} = -\frac{h_1^d(x)} {f_1^d(x)} = x^2 \,\, , 
\label{ssa} 
\eeq 
which is quite different from the simple $SU(6)$ values,  
\beqn 
\frac{h_1^u(x)} {f_1^u(x)} = \frac{2}{3}, \;\;\ 
\frac{h_1^d(x)} {f_1^d(x)} = - \frac{1}{3} \;\; . 
\eeqn 
We show our numerical result in Fig.11.   
${h_1^d(x)}/ {f_1^d(x)}$ and ${g_1^d(x)}/ {f_1^d(x)}$ are shown by the 
solid and dashed curves, and  
constraint from (\ref{ssa})  by  crosses.   
Our results are well consistent with the constraint (\ref{ssa}) from 
the single pion production.    
The QCD parametrization of ${g_1^d(x)} / {f_1^d(x)}$ is 
also shown by the circles \cite{Brodsky}.    
Note that calculated distribution functions never violate the Soffer's 
inequality in any cases.   
Rather singular behavior of $\frac{h_1^d(x)} {f_1^d(x)} = -1$ at $x=1$ is 
necessary to reproduce the observed large spin asymmetry of the  
$\pi^-$ production.  If we used the simple $SU(6)$ value  
${h_1^d(x)}/ {f_1^d(x)} = -1/3$, the resulting spin asymmetry would become 
about a half of the experimental value\cite{Exp_pi}.   
It is of great interest to investigate further approach 
to measure the $h_1(x)$ 
structure function, in particular its flavor dependence.

\begin{figure} 
\begin{center} 
\psfig{file=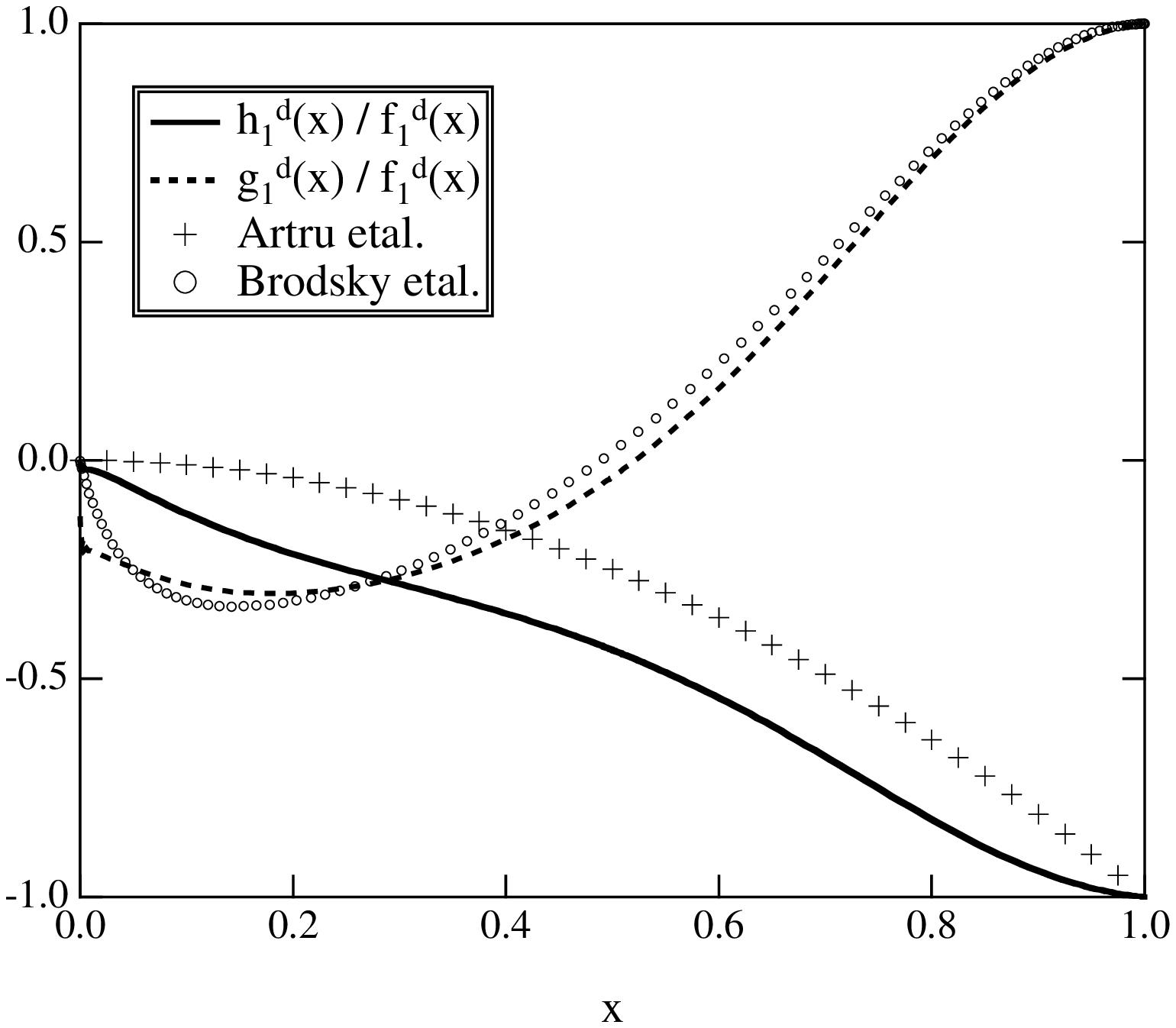,height=4in} 
\caption{Ratios of spin distribution functions $h_1^d(x) / f_1^d(x)$ and $g_1^d(x) / f_1^d(x)$\label{fig_artru}}    
\end{center} 
\vspace{-0.2cm} 
$h_1^d(x) / f_1^d(x)$ and $g_1^d(x) / f_1^d(x)$ are shown by the solid and 
dashed curves, respectively.   
Best fit parameterization for $h_1^d(x) / f_1^d(x)$  
extracted from the single spin asymmetry of the  
pion production in the polarized $pp$ collision\cite{Single} is also shown  
by the crosses.  The circles denote the parameterization of Brodsky  
{\etal} for $g_1^d(x) / f_1^d(x)$ \cite{Brodsky}. 
\end{figure} 

There are some difficulties to calculate the structure function within the 
effective quark model.   
Most serious one is use of the perturbative QCD below 1 GeV.    
To obtain better agreements with  the experimental data, the evolution from  
too small scale $\mu ^2 = 0.2 \sim 0.3 \GeV^2$ is needed.   
We show in Fig.12 $xf_1^u(x)$ obtained by LO and NLO evolution to  
clarify size of higher order corrections.   
Here, calculations with LO and NLO evolution are displayed by the  
solid and dashed curves, respectively.  
We use the same value of $\Lambda _ {QCD}$ to calculate both LO and NLO 
results for simplicity, 
though the LO and NLO values should be in general different as found in 
various QCD fits of the structure function\cite{Gluck}.  
It is found that difference is about  
$20 \%$ of the original result, which is not significant.   

\begin{figure} 
\begin{center} 
\psfig{file=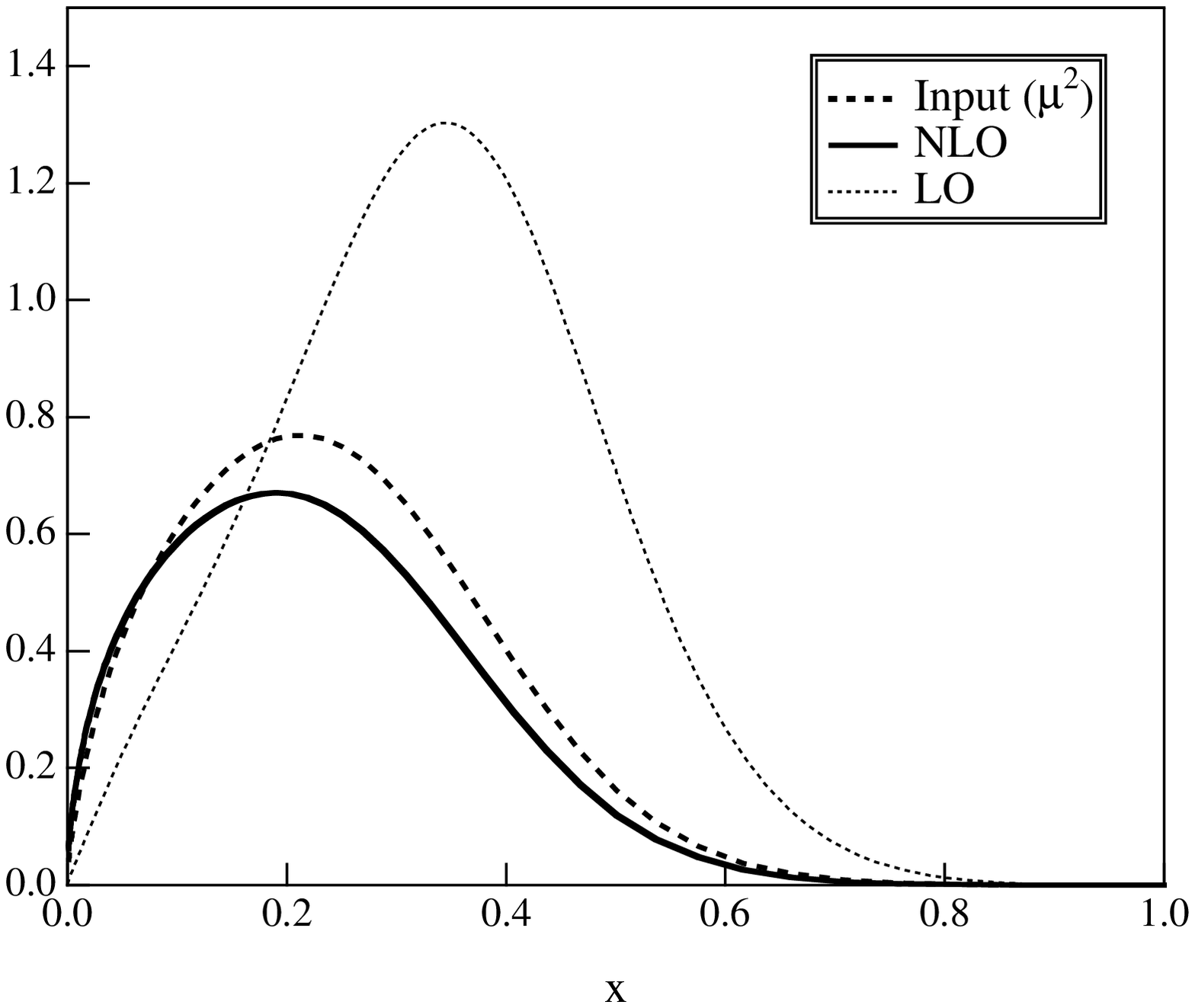,height=3in} 
\caption{Comparison of LO and NLO QCD evolution\label{fig_nlo}} 
\end{center} 
\vspace{-0.2cm}
Results for unpolarized  $u$-quark distribution functions  
with the NLO and LO $Q^2$ evolution are shown by the solid and dashed curves. 
\end{figure}

Whatever difference between  
the NLO and LO calculations is about 20$\%$,  validity of the 
perturbative QCD at such a scale is still questionable.   
It has recently been pointed out by Kulagin {\etal}\cite{Kulagin} that  
inclusion of the constituent quark structure improves this difficulty.     
They have shown that, if one takes into account the internal structure of the 
constituent quark, resulting distribution 
function becomes much softer.  Using such improved distributions as 
inputs of 
the $Q^2$ evolution at $1\GeV^2$, where the use of perturbative QCD can be 
justified, they have obtained remarkable agreements with the data.   
Our present work is extension of their work to the spin structure 
function.   
Although we have found that their model parameters gives  
too small values for the spin dependent properties like $g_A$,  
such efforts contribute to right direction to overcome this difficulty.

Our aim is to investigate the non-perturbative nucleon structure from studies  
of the quark distributions  measured in the high energy experiments.   
Obtained suppression of the $d$-quark tensor charge is a key to understand how 
the real nucleon structure differs from the simple bag model-like structure.   
Future experimental efforts will clarify spin-flavor structure of 
the nucleon.

\vspace{1cm} 
 
\ni 
{\bf Acknowledgments} 
 
K.S. would like to thank W.~Weise for reading of a manuscript and a  
collaboration 
of Ref.~\cite{Suzuki}, where corrections to the quark distributions from the  
Goldstone boson dressing are studied.   
This work is supported in part by the Alexander von Humboldt foundation.  

%
%
 
%
\new
{\bf Table Captions}

\ni
Table 1

The first moments of the helicity and transversity distribution functions  
(at $\mu^2$).    The axial charges $\Delta u$ and $\Delta d$ are shown in the
second and third columns, and corresponding tensor charges are given in the 
forth and fifth columns, respectively.  Results of with the bare CQ and dressed
CQ are displayed in the second and third raws.  Experimental data, lattice
calculations and QCD sum rule results are shown in the third, forth and fifth
raws, respectively.

\begin{center} 
 
{\bf Table 1}

\vspace{1cm} 
 
\begin{tabular}{l|c|c|c|c} 
\hline 
        & $\Delta u $   & $\Delta d $   & $\delta u $  & $\delta d$\\ \hline 
Bare    & 0.95          &$-0.30$        & 1.17         & $-0.26$   \\  
With CQ & 0.73          &$-0.25$        & 0.92         & $-0.15$   \\  
Exp\cite{Ellis}   & $0.83\pm 0.03$& $-0.43\pm0.03$&   $- $    &   $-$ \\ 
Lattice\cite{h1lattice} &  0.76         & $-0.35$     & 0.84         
 & $-0.23$   \\  
Sum rule\cite{QSR}    &   &   &  $1.33 \pm 0.53$      & $-0.04 \pm 0.02$ \\ 
\hline 
\end{tabular} 
 
\end{center} 
 
\end{document}